# Magnetic Field Dependent Piezoelectricity in Atomically Thin Co$_2$Te$_3$

*Solomon Demiss [†,‡], Alexey Karstev [¥,Φ]\*, Mithun Palit [Ψ], Prafull Pandey [Γ], G. P Das [†], Olu Emmanuel Femi [‡], Ajit K. Roy [§], Partha Kumbhakar [†]\*, Pulickel M. Ajayan [ξ], Chandra Sekhar Tiwary [†]\**

[†]Metallurgical and Materials Engineering, Indian Institute of Technology Kharagpur, Kharagpur-721302, India

[‡]Materials Science and Engineering, Jimma Institute of Technology, Jimma University, Jimma, Ethiopia

[¥]Stanford Institute for Materials and Energy Sciences, SLAC National Accelerator Laboratory, Menlo Park, CA-94025, USA

[Φ] Computing Center of Far Eastern Branch of the Russian Academy of Sciences (CC FEB RAS), Khabarovsk, Russian.

[Ψ] Defence Metallurgical Research Laboratory, Hyderabad, 500058, India

[Γ] Department of Materials Engineering, Indian Institute of Science, Bangalore 560012, India

[§]Materials and Manufacturing Directorate, Air Force Research Laboratory, Wright Patterson AFB, OH 45433-7718, United States

[ξ] Department of Materials Science and Nano Engineering, Rice University, Houston, TX 77005, USA

**ABSTRACT**

Two dimensional (2D) materials have received a surge in research interest due to their exciting range of properties. Here we show that 2D cobalt telluride ($Co_2Te_3$), successfully synthesized via liquid-phase exfoliation in an organic solvent, exhibits weak ferromagnetism and semiconducting behavior at room temperature. The magnetic field-dependent piezoelectric properties of 2D $Co_2Te_3$ sample show magneto-electric response of the material and a linear relationship between the output voltage and applied magnetic field. First-principles density functional theory (DFT) and ab initio molecular dynamics are used to explain these experimental results. Our work could pave the way for the development of 2D materials with coupled magnetism and piezoelectricity, leading to new applications in electromagnetics.

## 1. INTRODUCTION

In order to meet the growing demand for power supplies for microelectronic technology, energy harvesting technologies, wireless devices, and contactless sensors have developed in recent years. [1,2] Piezoelectric-based self-powered systems have created tremendous attention due to their ability to convert mechanical energy to electricity. Therefore, a piezoelectric based nanogenerator is well suited for microelectronics devices. Long term structural stability of these piezoelectric devices is questionable due to surface wear and tear generated due to mechanical forces. Hence, a contactless nanogenerator is a new kind of device where power will be devolved without mechanical force/pressure. Magnetic field-induced energy harvesting is one of the most promising ways to produce mechanical oscillation via magneto-mechano coupling, which in turn gets converted into electrical energy via mechano-electric coupling.[3] Contactless self-powered electronic systems such as sensors, biomedical equipment, and energy harvesters are essential components for next-generation electronic gadgets. All these magnetic-based energy harvesting devices need a large amount of field. Hence, there is a requirement for new materials that can produce energy in the small magnetic field.

Two-dimensional (2D) materials have fascinating physical, chemical, and mechanical properties, and they offer opportunities for electronics, optoelectronics, and energy storage applications.[4] The ultrathin 2D materials are more sensitive because of their flexible nature and therefore, they have been widely used for sensing devices. 2D materials also show emergence piezoelectric properties due to their unique characteristics such as crystal symmetry, surface defects, the capability of withstanding strain, doping etc.[5-8] On the particular compound of 2D Transition Metal Dichalcogenides (TMDCs) group, telluride-based materials such as bismuth tellurides, tin tellurides, antimony tellurides, tungsten tellurides, cobalt telluride, gallium telluride etc. are prevalent due to their unique properties and potential applications.[9-11] Recently, a small number of 2D telluride magnets have been obtained experimentally. When materials are taken to very low dimensions, the magnetic properties become more prominent.[12-14] Layer-dependent magnetism will provide more opportunities for their applications in sensing, memory devices, magnetic switching, spintronics etc.[15-17] Cobalt-based tellurides such as CoTe, $CoTe_2$ and $Co_2Te_3$, have been explored as a promising noble-metal-free material for different potential applications.[18-22] Amongst other properties, magnetic properties are also investigated for different structures of these tellurides.[20-22]

In the present work, we have successfully isolated 2D $Co_2Te_3$ via the liquid phase exfoliation method and investigate its magnetic field-dependent piezoelectric behavior. Spectroscopy and microscopy results confirm the formation of atomically thin $Co_2Te_3$ from their bulk counterpart.

The semiconducting nature was also confirmed from the optical absorption measurement. Besides, we have studied the magnetic property of 2D $Co_2Te_3$ at room temperature. We have systemically investigated the generation of the output voltage by piezoelectric behavior under a variable magnetic field. Theoretical studies based on density functional theory (DFT) and ab initio molecular dynamics confirm the magnetic and piezo-magnetic properties of the ultrathin $Co_2Te_3$, which are in good agreement with the experimental observation.

## 2. EXPERIMENTAL SECTION

Bulk $Co_2Te_3$ material was fabricated using the vacuum induction melting method. The stoichiometric composition of the sample was selected based on Co-Te Binary Alloy Phase Diagram. High purity elements (99.99% purity) purchased from Sigma Aldrich were used for preparing the compound containing 25 wt % of Co, and 75 wt % of Te. This composition range is selected because the Co2Te3 phase is a more stable 60 at % tellurium composition within Co-Te system. The alloys were prepared by melting the constituent element at a temperature of 1050°C in a quartz tube using a Vacuum induction melting technology under argon atmosphere and allowed to cool naturally under argon. High vacuum condition of $1\times10^{-5}$ mbar was maintained inside the melting chamber to ensure high purity and prevent from oxidation. The bulk samples with 10cm × 8cm × 1cm were made and cleaned, and polished for further characterization. 2D $Co_2Te_3$ nanosheets were fabricated by sonication-assisted liquid exfoliation by mixing the mechanically grinded 50 mg of $Co_2Te_3$ powder with 150 mL of IPA solvent in an ultrasonic vibrator for 6 hours at room temperature to obtain a suspension of 2D $Co_2Te_3$ (Figure 1a). After centrifugal treatment, few-layer $Co_2Te_3$ flakes were dispersed in the solvent.

The morphology and composition of the samples were observed using JEOL (JED 2300) scanning electron microscope (SEM) under a high vacuum at 20kV and 7.475nA. The crystal structure and purity of bulk and 2D $Co_2Te_3$ was examined using Single Crystal X-ray Diffractometer (XRD), a Bruker D8 advance diffractometer with Cu-Kα (λ=1.5406 Å) radiation (2θ=10º-110º) with scan speed of 0.3 for 27 min. The diffraction peaks were analyzed and indexed using Panalytical Xpert high score plus software and Pearson's Crystal Database Software. Room temperature Raman spectroscopy was performed using a WITec UHTS Raman Spectrometer (WITec, UHTS 300 VIS, Germany) at a laser excitation wavelength of λ=532 nm. A Shimadzu 2450 UV-Visible spectrophotometer was employed to obtain UV-Visible absorption measurements. The magnetization measurements were carried out using VSM instrument at room temperature by applying a magnetic field from -15 to +15 kOe. Powder Samples weight of bulk $Co_2Te_3$ (98.23 mg) and 2D $Co_2Te_3$ (56.81mg) were used.

## 3. RESULTS AND DISCUSSIONS

**Figure 1a** depicts the synthesizing process of 2D $Co_2Te_3$ from its bulk counterpart. The as fabricated bulk sample was crushed and dispersed in Isopropanol Alcohol (IPA) (see the experimental section for more details). **Figure 1b** shows the X-ray diffraction (XRD) patterns of the bulk and exfoliated $Co_2Te_3$. The peaks are well matching with the hexagonal crystal structure and the lattice parameters *a= b*= 3.8427 Å and *c*= 5.3794 Å and *α=β*=90°; *γ*=120° and in a good agreement with reported literature.[23] For bulk $Co_2Te_3$ the maximum of profile intensity occurring at the (101), (002), (102), (110), (201), (103), and (202) diffraction peaks (see **Figure 1b** bottom). As noted in **Figure 1b**, peaks of (100), (101), (002), (102), (110), and (103) were observed in the 2D $Co_2Te_3$. Among these peaks, the (100), (101) and (002) diffraction peaks are dominant, which suggests that these planes were exfoliated in excess during the exfoliation method compared to other planes from bulk samples. The different diffraction peaks at (100) for 2D and (101) for bulk shows the preferential plane orientation of each sample. Raman spectroscopy was used to ensure the 2D characteristics of $Co_2Te_3$. The Raman peaks of exfoliated $Co_2Te_3$ (see **Figure 1c**) were observed at 53, 180, 320, and 377 cm$^{-1}$ which are consistent with the in-plane and out-of-plane vibrational mode of Transitional Metal Tellurides (TMTs).[24] XPS measurement was carried out to identify the elemental composition and oxidation states of the exfoliated $Co_2Te_3$. The survey spectra (**Figure 1d**) confirm the presence of Co and Te. **Figure 1e** shows the high-resolution spectra with spin-orbit interaction of Co $2p_{3/2}$ and Co $2p_{1/2}$ at 780.5, 786.4 and 796.4, 801.4 eV, respectively. The energy difference of 15.7 eV confirms the presence $Co^{3+}$ ion state.[25] **Figure 1f** presents the high-resolution spectrum of Te 3d. The satellites at 573.8 eV and 584.2 eV correspond to Te 3d5/2 and Te 3d3/2 peaks of the Te2- ionic state, while the major peaks at 576.3eV 586.7 eV arise due to the $Te^{4-}$ species.[26]

**Figure S1-S3** illustrates the SEM image of bulk and exfoliated $Co_2Te_3$. The SEM result strongly suggests that the material is exfoliated into individual layers. The morphology of 2D $Co_2Te_3$ is also characterized using bright-field (BF) TEM measurement and shows ultrathin flakes with lateral dimensions of ~200-400 nm (**Figure 2a**). Figure 2b presents the High-resolution BF TEM image of exfoliated Co2Te3 and the image clearly illustrates the lattice pattern. We have calculated the lattice spacing, and it is found to be ~0.21 nm, which agrees with (101) plane of $Co_2Te_3$ (**Figure 2c**). Elemental analysis was performed using Energy Dispersive X-Ray (EDX) Spectroscopy that shows the obtained material to consist only of Co and Te elements with nearly the same atomic percentage ratio of Te and Co, consistent with the expected stoichiometric ratio (**Figure 2d**). Atomic force microscopy (AFM) image confirms the formation of the atomically thin sheet (**Figure 2e**). It indicates a sheet-like structure with an average thickness of ~2-3 nm (inset of **Figure 2e**).

The magnetic behaviour of bulk and 2D $Co_2Te_3$ samples was measured using a vibrating sample magnetometer (VSM) at room temperature. 2D $Co_2Te_3$ has a weak ferromagnetic ground state, as one could see from the M-H loops (**Figure 2f**) at 300K with an excellent degree of saturated magnetization. The magnetic hysteresis curve (Figure 2f) has estimated Magnetic susceptibility, Magnetic saturation, and Coercivity values for bulk and 2D $Co_2Te_3$. The estimated values of Magnetic susceptibility are ~1.204 $\times 10^{-5}$ and 1.867 $\times 10^{-5}$ emu/g Oe for bulk and 2D samples at room temperature. We have also calculated the coercivity, and it is found to be 53.37 and 55.70 Oe, for bulk and 2D, respectively. We also estimate the retentivity for both bulk and 2D materials and it was increased from 0.0643$\times 10^{-3}$ to 1.040$\times 10^{-3}$ emu/g, from bulk to 2D, respectively. As shown in the Magnetic saturation was increased from 0.06 $\times 10^{-3}$ to 0.09 $\times 10^{-3}$ emu/g as thickness decreased (inset of **Figure 2f**). The magnetic saturation of exfoliated $Co_2Te_3$ sample increased by ~50% as compared to its bulk counterpart. Therefore, 2D $Co_2Te_3$ is a weak ferromagnetic semiconductor material that can potentially apply to magnetic sensors, micro electrochemical systems, etc.

The optical absorption property of exfoliated $Co_2Te_3$ was analyzed using UV-Vis Spectroscopy in the energy range of 1.5 to 3.5 eV. **Figure S4** shows the absorbance spectrum of the 2D $Co_2Te_3$ dispersed in IPA. The observed absorption bands indicate that the absorption coefficient of 2D $Co_2Te_3$ is higher in the visible region of the spectrum. The band edge position of bulk and 2D $Co_2Te_3$ is estimated from the second derivative of UV-Vis spectrum (inset of **Figure S4**) The energy band gap value was determined by extrapolating the straight-line portion of the curves to zero absorption coefficient value. We found that the bandgap value was $E_g$ = 1.97 eV and ~2.6 eV for bulk and exfoliated samples.

Piezoelectric Effect is the ability of centrosymmetric materials to generate an electric polarization response to applied mechanical strain gradient or the mechanical response under an electric field gradient.[27] **Figure 3a** shows a schematic of the fabrication process of the PNG cell, illustrating the ultrathin $Co_2Te_3$ based piezoelectric nanogenerator under applied load. **Figure 3b** shows the output voltage generated from the fabricated device as a function of time. For 2D materials, the output voltage increases as applied load increases (0.1N to 0.4N), and it is due to the symmetry deformation of the structure. When 2D materials are deformed, the output power density and conversion efficiency increase significantly as the sample thickness reduces from micro to nanoscale. The as-fabricated nanogenerator shows a very fast response time of ~5 ms as shown in **Figure 3c**. In **Figure 3d,** we present the increment of an output voltage of the fabricated device as a function of the applied load. As the applied load increases from 0.1 to 0.4 N, the output voltage increases rapidly from ~2 to 4 V, indicating that our exfoliated $Co_2Te_3$ can be used as the functional material for energy harvesting or self-powered acceleration sensor through appropriate electrical design. The output voltage is measured as a function of external load resistance ($R_L$), keeping a constant applied force of 0.3N. The output voltage increased with increase the value of $R_L$ and reached at maximum voltage. The calculated power density [28] is found to be ~1.17 mW/m² (see supporting information for more details)

We demonstrate the magnetic force-induced piezoresponse of the 2D $Co_2Te_3$. The magnetic force-derived piezo potential is essential in magnetic induction application. For contactless devices, we use the magnetic field as a source of strain in the material. **Figure 4a** depicts the working mode of the piezoelectric nanogenerator under the applied magnetic field. The strain changes into the $Co_2Te_3$ based device due to the movement of magnets. When magnets are moving back and forth, lines of force also change. Therefore, piezoelectric potential varied due to the magnetic field by generating strain inside the material. **Figure 4b** shows the output voltage as a function of time with and without applied magnetic field. The output voltage increased when magnetic force is applied indicates the applied magnetic field affects the piezoelectric property of the 2D $Co_2Te_3$ nanogenerator. **Figure 4c** shows the enlarged view of one complete cycle in the movement of a magnet. Actually, the material reported here is unique. It is ferromagnetic as well as piezoelectric. Usually to create a magneto-electric effect or in so-called multiferroics, a heterostructure or composite is made of a ferromagnetic phase and a piezoelectric material. Fortunately, in this case, both are available together. When magnets are approaching the material, it interacts with magnetic domains and produces strain in magnetostriction materials. This strain produces stress in the material, adding up to the externally applied load and producing more voltage. The generated voltage signal approaches ~0.2 V on average, which is an advantage in the application of magnetic sensing. In **Figure 4d,** we illustrate the output voltage of the cell as a function of a magnetic field. All experimental points can be fitted using a straight line, which indicates a well response nature of the device. We have also calculated the responsivity (dV/dB) of the piezo-magnetic nanogenerator and it is found to be ~4.1 m V/T. In **Figure 4e**, we demonstrate the output voltage of the piezo-magnetic nanogenerator as a function of the distance between the magnets and the device. When we reduce the distance to the cell, the piezo potential under the magnetic field increases this is because as we approach to the cell the magnetic lines of force change and magnetic field increase. Due to this variation, we observed that as the magnetic field increases, i.e., approaches to the cell, the magnetic lines of forces increase and increase output voltage from ~0.12 V to 0.19 V.

**Theoretical quantification of magneto-electric response 2D-$Co_2Te_3$**

In order to complement our experimental results, we have carried out first-principles DFT calculation on 2D-$Co_2Te_3$ monolayer (4×4 supercell 80-atoms supercell with 20Å vacuum) using the PAW method implemented in the VASP code (upper panel of **Figure 5, Figure S5 and S6,** see supporting information for more details). The unit cell corresponds to non-centrosymmetric space group #187 of $Co_2Te_3$. We have used PBE exchange-correlation potential within the GGA+U framework (choosing the Hubbard $U_{eff}$ = 3eV for 2D-$Co_2Te_3$ compound) and DFT-D3 dispersion correction into account the weak van der Waals interlayer bonding between 2D-$Co_2Te_3$ layers. Furthermore, to study the structural as well as temperature stability of the system, we have carried out *ab initio* MD simulation for the Monolayer $Co_2Te_3$ system with canonical ensemble using Nose-Hoover thermostat (NVT) (details included in supporting information)

The results for relaxed ground state lattice parameter ($a_0$=3.82 Å) and bond lengths ($d_{Co-Co}$=2.65 Å and $d_{Co-Te}$=2.59 Å) of monolayer (1L)-$Co_2Te_3$ structure are in excellent agreement with observed values. The self-consistent charge density distribution (**Figure 5**) indicates the formation of the dipole moment in the deformed material. We have also plotted the electronics localization function (ELF) [30] (see **Figure S7** in Supporting Information) showing how the electrons tend to be localized near the chalcogen atom, although the relatively small electronegativity difference (~0.22) between Co and Te leads to a polar-covalent character rather than ionic character. The estimated magnitudes of the charge transfers from the Co atoms to the outer and inner Te atoms are about 0.09e and 0.02e, respectively. This charge transfer difference leads to the net electric field directed from the outer Te to Co. This is in sharp contrast to that of nonpolar-covalent materials like graphene, where electrons are usually located right in the middle between carbon atoms. In order to reiterate the nature of chemical bonding in 1L-$Co_2Te_3$, we have plotted its crystal orbital Hamilton population (COHP) and the projected COHP using projection of the PAW basis. The observed *p-d* hybridization for $Co_2Te_3$ monolayer is shown in **Figure S8** of the supporting information.

To further examine the chemical stability of the 2D-$Co_2Te_3$, we have computed the formation enthalpy per formula unit, $H_f$ = -0.62 eV/$Co_2Te_3$. The interlayer binding energy ($E_b$) is estimated to be 120 meV/Å$^2$, which is 3.5 times higher than that for a mechanically exfoliated layer. [29] *Ab initio* molecular dynamics (AIMD) simulations confirm the structural stability up to at least ~700K, indicating defect bonds formation at higher temperature T=1000-1350K.

Several magnetic configurations for 1L-$Co_2Te_3$ viz. nonmagnetic, ferromagnetic, ferrimagnetic, and 5 different antiferromagnetic configurations, have been analyzed using the supercell shown in **Figure 5** (lower panel of) The ground state is found to be ferromagnetically ordered with the local magnetic moment on Co $\mu_{Co}$=0.77$\mu_B$ and the estimated Curie temperature $T_C \approx \Sigma I_{ij}/3k_B$=180 K. The resulting magneto-crystalline anisotropy $E_{MAE}$ = 1.68 meV per Co-atom corresponds to the easy *z*-axis.

## 4. CONCLUSIONS

In conclusion, 2D $Co_2Te_3$ is successfully exfoliated from the as-fabricated bulk crystal $Co_2Te_3$ via the liquid-phase exfoliation method. The exfoliated $Co_2Te_3$ are found to be a weakly ferromagnetic semiconductor material. 2D $Co_2Te_3$ also used as a nanogenerator using piezoelectric effect, with an output voltage of ~3.6 V. In addition, the magneto-electric property was also investigated under different magnetic fields at room temperature. A microscopic understanding of the origin of the magneto-electric response has been obtained using state-of-the-art density functional calculations that reiterate the magneto-crystalline anisotropy of this 2D monolayer system. The existence of such magneto-electric response in a single material is highly desirable in nanoscale device applications.

## ASSOCIATED CONTENT

**Supporting Information:** Computational details, Optical property, SEM, Piezoelectric calculation,

## AUTHOR INFORMATION

**Corresponding Author**

Alexey Karstev

Stanford Institute for Materials and Energy Sciences, SLAC National Accelerator Laboratory, Menlo Park, CA-94025, USA

Computing Center of Far Eastern Branch of the Russian Academy of Sciences (CC FEB RAS), Khabarovsk, Russian.

E-mail: karec1@gmail.com

Partha Kumbhakar* and Chandra S. Tiwary*

Metallurgical and Materials Engineering, Indian Institute of Technology Kharagpur, Kharagpur-721302, India

E-mail: parthakumbhakar2@gmail.com

E-mail: chandra.tiwary@metal.iitkgp.ac.in

**ACKNOWLEDGMENT**

Solomon Demiss (S.D) acknowledges Jimma Institute of Technology/Jimma University and Ministry of Science and Higher Education Ethiopia for their funding support. We also thank Indian institute of technology (IITKGP) for laboratory facility support. Simulation work by Alexey Kartsev (A.K) was supported by the US Department of Energy (DOE), Office of Basic Energy Sciences, Materials Sciences and Engineering Division, under contract number DE-AC02-76SF00515 through the Center for Non-Perturbative Studies of Functional Materials (NPNEQ)". We would like to acknowledge the Shared Facility Center resources 'Data Center of FEB RAS' (Khabarovsk) (citation Sorokin, A. A., Makogonov, S. V., & Korolev, S. P. (2017). The information infrastructure for collective scientific work in the Far East of Russia. Scientific and Technical Information Processing, 44(4), 302-304.) and Joint Supercomputer Center of the Russian Academy of Sciences (JSCC RAS) where the data processing was performed. A.K. would like to thank Chaitanya Das, Pemmaraju and Tadashi Ogitsu, who significantly contributed to the quality of this paper with valuable and well-aimed comments.

# REFERENCES

(1) Cui, N.; Wu, W.; Zhao, Y.; Bai, S.; Meng, L.; Qin, Y.; Wang, Z. L. Magnetic Force Driven Nanogenerators as a Noncontact Energy Harvester and Sensor. *Nano Lett.* **2012**, *12* (7), 3701–3705. https://doi.org/10.1021/nl301490q.

(2) Liu, Y.; Guo, J.; Yu, A.; Zhang, Y.; Kou, J.; Zhang, K.; Wen, R.; Zhang, Y.; Zhai, J.; Wang, Z. L. Magnetic-Induced-Piezopotential Gated MoS$_2$ Field-Effect Transistor at Room Temperature. *Advanced Materials* **2018**, *30* (8), 1704524. https://doi.org/https://doi.org/10.1002/adma.201704524.

(3) Ryu, J.; Kang, J.-E.; Zhou, Y.; Choi, S.-Y.; Yoon, W.-H.; Park, D.-S.; Choi, J.-J.; Hahn, B.-D.; Ahn, C.-W.; Kim, J.-W.; Kim, Y.-D.; Priya, S.; Lee, S. Y.; Jeong, S.; Jeong, D.-Y. Ubiquitous Magneto-Mechano-Electric Generator. *Energy Environ. Sci.* **2015**, *8* (8), 2402–2408. https://doi.org/10.1039/C5EE00414D.

(4) Cui, C.; Xue, F.; Hu, W.-J.; Li, L.-J. Two-Dimensional Materials with Piezoelectric and Ferroelectric Functionalities. *npj 2D Materials and Applications* **2018**, *2* (1), 1–14. https://doi.org/10.1038/s41699-018-0063-5.

(5) Khan, H.; Mahmood, N.; Zavabeti, A.; Elbourne, A.; Rahman, M. A.; Zhang, B. Y.; Krishnamurthi, V.; Atkin, P.; Ghasemian, M. B.; Yang, J.; Zheng, G.; Ravindran, A. R.; Walia, S.; Wang, L.; Russo, S. P.; Daeneke, T.; Li, Y.; Kalantar-Zadeh, K. Liquid Metal-Based Synthesis of High Performance Monolayer SnS Piezoelectric Nanogenerators. *Nature Communications* **2020**, *11* (1), 3449. https://doi.org/10.1038/s41467-020-17296-0.

(6) Ong, M. T.; Reed, E. J. Engineered Piezoelectricity in Graphene. *ACS Nano* **2012**, *6* (2), 1387–1394. https://doi.org/10.1021/nn204198g.

(7) Duerloo, K.-A. N.; Ong, M. T.; Reed, E. J. Intrinsic Piezoelectricity in Two-Dimensional Materials. *J. Phys. Chem. Lett.* **2012**, *3* (19), 2871–2876. https://doi.org/10.1021/jz3012436.

(8) Apte, A.; Mozaffari, K.; Samghabadi, F. S.; Hachtel, J. A.; Chang, L.; Susarla, S.; Idrobo, J. C.; Moore, D. C.; Glavin, N. R.; Litvinov, D.; Sharma, P.; Puthirath, A. B.; Ajayan, P. M. 2D Electrets of Ultrathin MoO2 with Apparent Piezoelectricity. *Advanced Materials* **2020**, *32* (24), 2000006. https://doi.org/https://doi.org/10.1002/adma.202000006.

(9) Li, S.; Toprak, M. S.; Soliman, H. M. A.; Zhou, J.; Muhammed, M.; Platzek, D.; Müller, E. Fabrication of Nanostructured Thermoelectric Bismuth Telluride Thick Films by

Electrochemical Deposition. *Chem. Mater.* **2006**, *18* (16), 3627–3633. https://doi.org/10.1021/cm060171o.

(10) Duerloo, K.-A. N.; Ong, M. T.; Reed, E. J. Intrinsic Piezoelectricity in Two-Dimensional Materials. *J. Phys. Chem. Lett.* **2012**, *3* (19), 2871–2876. https://doi.org/10.1021/jz3012436.

(11) Kang, S.; Jeon, S.; Kim, S.; Seol, D.; Yang, H.; Lee, J.; Kim, Y. Tunable Out-of-Plane Piezoelectricity in Thin-Layered MoTe$_2$ by Surface Corrugation-Mediated Flexoelectricity. *ACS Appl. Mater. Interfaces* **2018**, *10* (32), 27424–27431. https://doi.org/10.1021/acsami.8b06325.

(12) Huang, B.; Clark, G.; Navarro-Moratalla, E.; Klein, D. R.; Cheng, R.; Seyler, K. L.; Zhong, D.; Schmidgall, E.; McGuire, M. A.; Cobden, D. H.; Yao, W.; Xiao, D.; Jarillo-Herrero, P.; Xu, X. Layer-Dependent Ferromagnetism in a van Der Waals Crystal down to the Monolayer Limit. *Nature* **2017**, *546* (7657), 270–273. https://doi.org/10.1038/nature22391.

(13) Fei, Z.; Huang, B.; Malinowski, P.; Wang, W.; Song, T.; Sanchez, J.; Yao, W.; Xiao, D.; Zhu, X.; May, A. F.; Wu, W.; Cobden, D. H.; Chu, J.-H.; Xu, X. Two-Dimensional Itinerant Ferromagnetism in Atomically Thin Fe$_3$GeTe$_2$. *Nature Mater* **2018**, *17* (9), 778–782. https://doi.org/10.1038/s41563-018-0149-7.

(14) Deng, Y.; Yu, Y.; Song, Y.; Zhang, J.; Wang, N. Z.; Sun, Z.; Yi, Y.; Wu, Y. Z.; Wu, S.; Zhu, J.; Wang, J.; Chen, X. H.; Zhang, Y. Gate-Tunable Room-Temperature Ferromagnetism in Two-Dimensional Fe$_3$GeTe$_2$. *Nature* **2018**, *563* (7729), 94–99. https://doi.org/10.1038/s41586-018-0626-9.

(15) Han, W.; Kawakami, R. K.; Gmitra, M.; Fabian, J. Graphene Spintronics. *Nature Nanotechnology* **2014**, *9* (10), 794–807. https://doi.org/10.1038/nnano.2014.214.

(16) Fei, Z.; Zhao, W.; Palomaki, T. A.; Sun, B.; Miller, M. K.; Zhao, Z.; Yan, J.; Xu, X.; Cobden, D. H. Ferroelectric Switching of a Two-Dimensional Metal. *Nature* **2018**, *560* (7718), 336–339. https://doi.org/10.1038/s41586-018-0336-3.

(17) Liu, H.; Wang, X.; Wu, J.; Chen, Y.; Wan, J.; Wen, R.; Yang, J.; Liu, Y.; Song, Z.; Xie, L. Vapor Deposition of Magnetic Van der Waals NiI$_2$ Crystals. *ACS Nano* **2020**, *14*, 10544−10551. https://dx.doi.org/10.1021/acsnano.0c04499.

(18) Gao, Q.; Huang, C.-Q.; Ju, Y.-M.; Gao, M.-R.; Liu, J.-W.; An, D.; Cui, C.-H.; Zheng, Y.-

R.; Li, W.-X.; Yu, S.-H. Phase-Selective Syntheses of Cobalt Telluride Nanofleeces for Efficient Oxygen Evolution Catalysts. *Angewandte Chemie International Edition* **2017**, *56* (27), 7769–7773. https://doi.org/https://doi.org/10.1002/anie.201701998.

(19) Liu, M.; Lu, X.; Guo, C.; Wang, Z.; Li, Y.; Lin, Y.; Zhou, Y.; Wang, S.; Zhang, J. Architecting a Mesoporous N-Doped Graphitic Carbon Framework Encapsulating CoTe2 as an Efficient Oxygen Evolution Electrocatalyst. *ACS Appl. Mater. Interfaces* **2017**, *9* (41), 36146–36153. https://doi.org/10.1021/acsami.7b09897.

(20) Dahal, B. R.; Dulal, R. P.; Pegg, I. L.; Philip, J. Electrical Transport and Magnetic Properties of Cobalt Telluride Nanostructures. *Journal of Vacuum Science & Technology B* **2016**, *34* (5), 051801. https://doi.org/10.1116/1.4959576.

(21) Lei, Y.-X.; Miao, N.-X.; Zhou, J.-P.; Hassan, Q. U.; Wang, J.-Z. Novel Magnetic Properties of CoTe Nanorods and Diversified $CoTe_2$ Nanostructures Obtained at Different NaOH Concentrations. *Science and Technology of Advanced Materials* **2017**, *18* (1), 325–333. https://doi.org/10.1080/14686996.2017.1317218.

(22) Muhler, M.; Bensch, W.; Schur, M. Preparation, Crystal Structures, Experimental and Theoretical Electronic Band Structures of Cobalt Tellurides in the Composition Range. *J. Phys.: Condens. Matter* **1998**, *10* (13), 2947–2962. https://doi.org/10.1088/0953-8984/10/13/012.

(23) Zhang, Z.; Hines, W. A.; Budnick, J. I.; Perry, D. M.; Wells, B. O. Direct Evidence for the Source of Reported Magnetic Behavior in "CoTe." *AIP Advances* **2017**, *7* (12), 125322. https://doi.org/10.1063/1.4997161.

(24) Zhou, J.; Liu, F.; Lin, J.; Huang, X.; Xia, J.; Zhang, B.; Zeng, Q.; Wang, H.; Zhu, C.; Niu, L.; Wang, X.; Fu, W.; Yu, P.; Chang, T.-R.; Hsu, C.-H.; Wu, D.; Jeng, H.-T.; Huang, Y.; Lin, H.; Shen, Z.; Yang, C.; Lu, L.; Suenaga, K.; Zhou, W.; Pantelides, S. T.; Liu, G.; Liu, Z. Large-Area and High-Quality 2D Transition Metal Telluride. *Advanced Materials* **2017**, *29* (3), 1603471. https://doi.org/https://doi.org/10.1002/adma.201603471.

(25) Liu, M.; Lu, X.; Guo, C.; Wang, Z.; Li, Y.; Lin, Y.; Zhou, Y.; Wang, S.; Zhang, J. Architecting a Mesoporous N-Doped Graphitic Carbon Framework Encapsulating $CoTe_2$ as an Efficient Oxygen Evolution Electrocatalyst. *ACS Appl. Mater. Interfaces* **2017**, *9* (41), 36146–36153. https://doi.org/10.1021/acsami.7b09897.

(26) Nardo, S. D.; Lozzi, L.; Passacantando, M.; Picozzi, P.; Santucci, S. Growth of Te Thin

Films Deposited at Room Temperature on the Si(100)2 × 1 Surface. *Journal of Electron Spectroscopy and Related Phenomena* **1995**, *71* (1), 39–45. https://doi.org/10.1016/0368-2048(94)02249-6.

(27) Jiang, X.; Huang, W.; Zhang, S. Flexoelectric Nano-Generator: Materials, Structures and Devices. *Nano Energy* **2013**, *2* (6), 1079–1092. doi.org/10.1016/j.nanoen.2013.09.001.

(28) Ghosh, S. K.; Adhikary, P.; Jana, S.; Biswas, A.; Sencadas, V.; Gupta, S. D.; Tudu, B.; Mandal, D. Electrospun Gelatin Nanofiber Based Self-Powered Bio-e-Skin for Health Care Monitoring. *Nano Energy* **2017**, *36*, 166–175. doi.org/10.1016/j.nanoen.2017.04.028.

(29) Fang, Z.; Li, X.; Shi, W.; Li, Z.; Guo, Y.; Chen, Q.; Peng, L.; Wei, X. Interlayer Binding Energy of Hexagonal MoS$_2$ as Determined by an *In Situ* Peeling-to-Fracture Method. *J. Phys. Chem. C* **2020**, *124* (42), 23419–23425. https://doi.org/10.1021/acs.jpcc.0c06828.

(30) Savin, A.; Nesper, R.; Wengert, S.; Fässler, T. F. ELF: The Electron Localization Function. *Angew. Chem. Int. Ed. Engl.* **1997**, *36* (17), 1808–1832. doi.org/10.1002/anie.199718081.

**Figures**

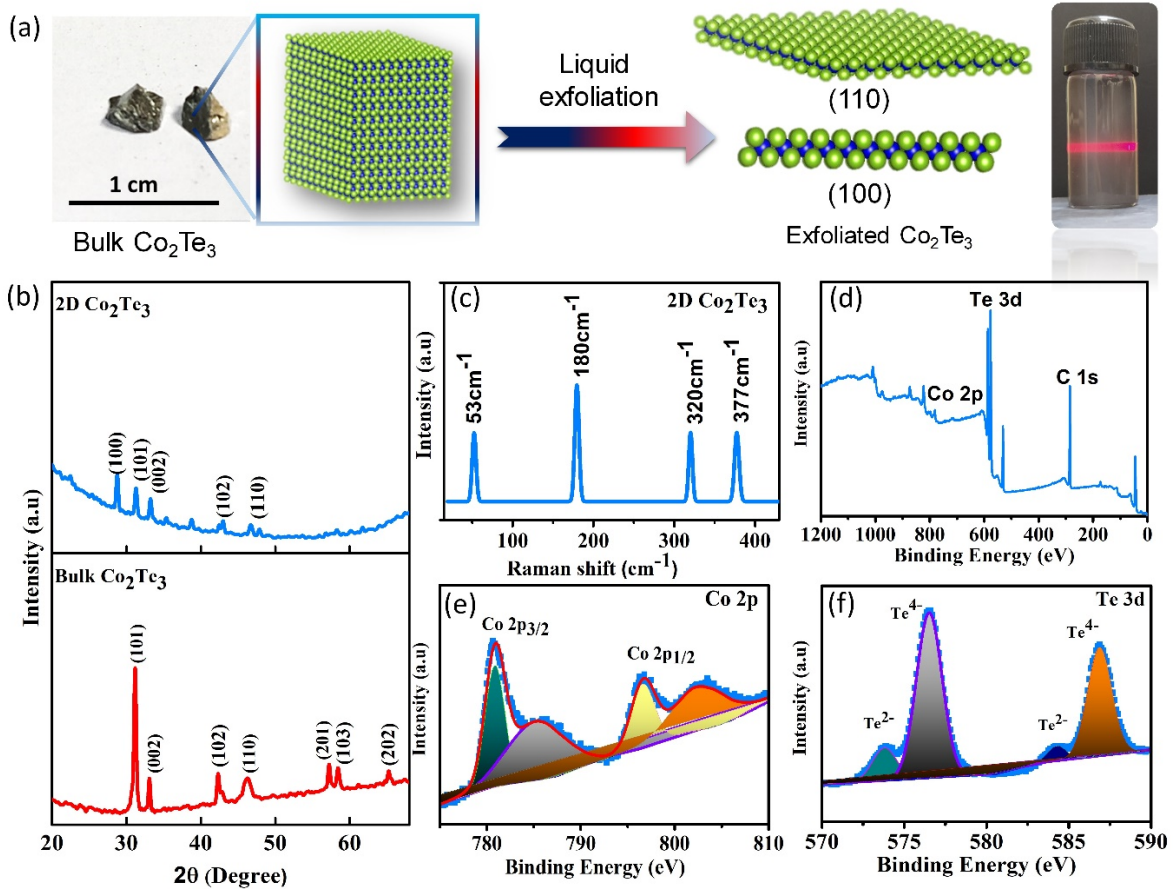

**Figure 1.** Schematic illustration of exfoliation process. Digital photograph of a Co$_2$Te$_3$ crystal and dispersion of 2D Co$_2$Te$_3$ in IPA solvent. Crystal structure of Bulk cobalt (III) and monolayer Co$_2$Te$_3$ made by VESTA software from X-ray diffraction results. (b) XRD pattern of bulk and 2D Co$_2$Te$_3$ from 2θ = 20 to 70°. (c) Raman analysis of the synthesized Co$_2$Te$_3$ measured at room temperature under excitation of λ = 532 nm. (d) Full XPS survey of exfoliated sample. High resolution XPS spectra corresponding to Co 2p (e) and (f) Te 3d of 2D Co$_2$Te$_3$.

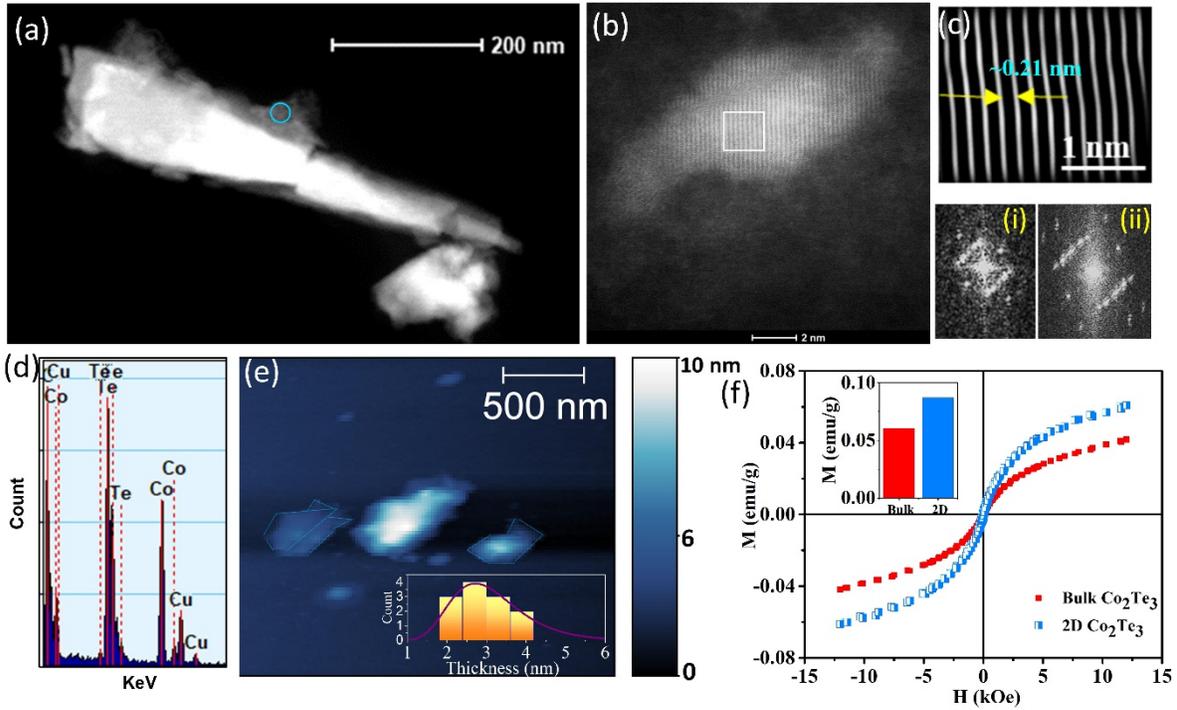

**Figure 2** (a) TEM images of exfoliated $Co_2Te_3$ show a 2D sheets. (b) HRTEM image of exfoliated sample. (c) Lattice spacing of exfoliated sample. Lower inset shoes the FFT patterns. (d) EDX pattern of exfoliated $Co_2Te_3$. (e) AFM images showing ultrathin nanosheets. Inset shows the bar chart presentation of thickness. (f) Magnetic property measurement of the exfoliated $Co_2Te_3$ showing M–H curves of 2D and bulk $Co_2Te_3$ at 300K in emu/g ($\times 10^{-3}$). Inset shows bar chart representation of magnetic saturation.

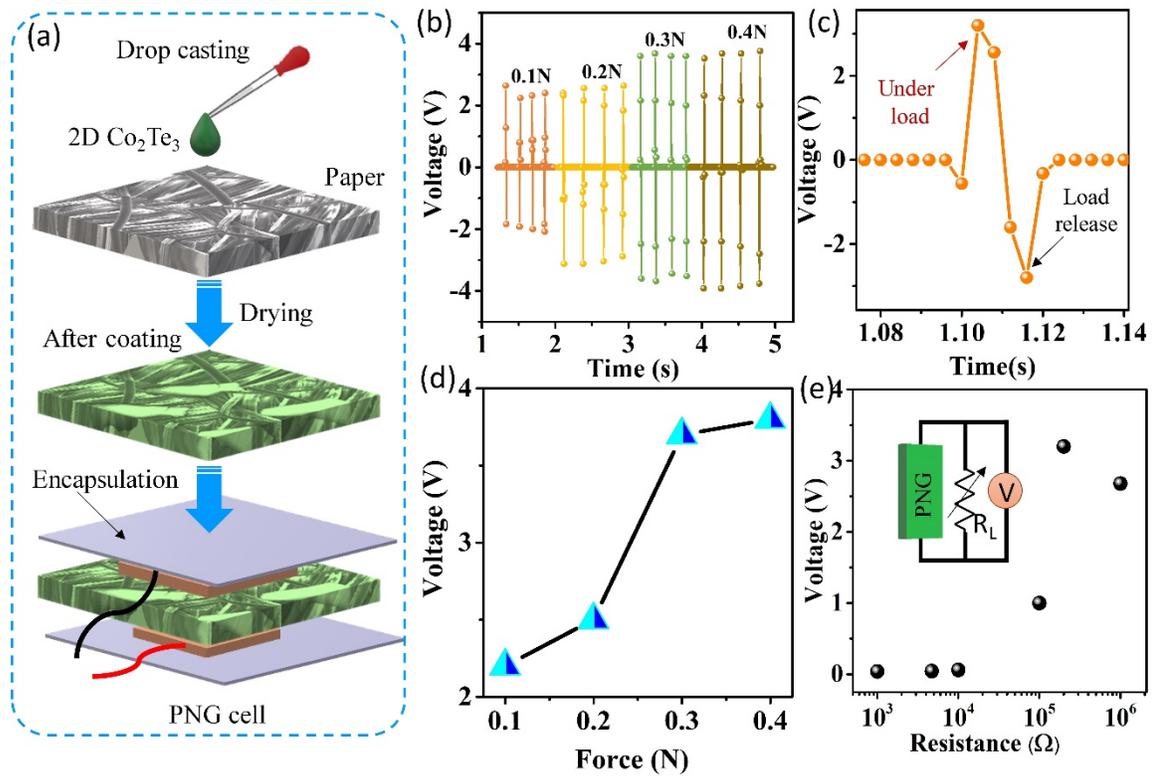

**Figure 3**: (a) Schematic diagram of fabrication method of the PNG cell. (b) Open circuit voltage response as a function of applied load. (c) enlarged views of a single pulse (d) the variation of output voltage as a function of applied load (e) Dependence of output voltage across variable resistances. Inset showing the circuit diagram

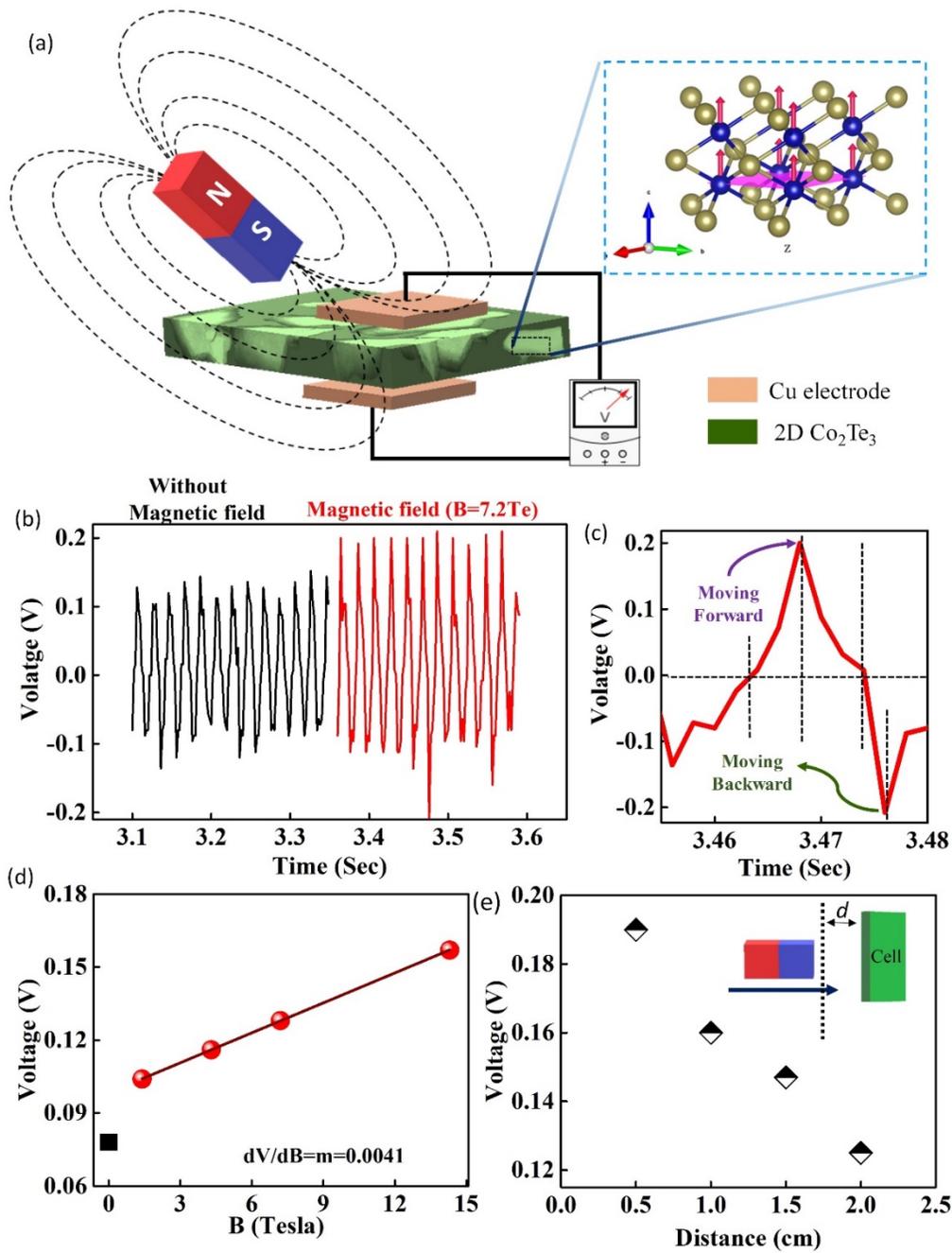

**Figure 4.** (a) Schematic diagram of the magnetic force assisted Co$_2$Te$_3$ based piezoelectric nanogenerator. (b) Output voltage with and without magnetic field. (c) the enlarged view of one full cycle. (d) the output voltage as a function magnet field. (e) The output voltage of the PNG as a function of the distance between the magnets and PNG.

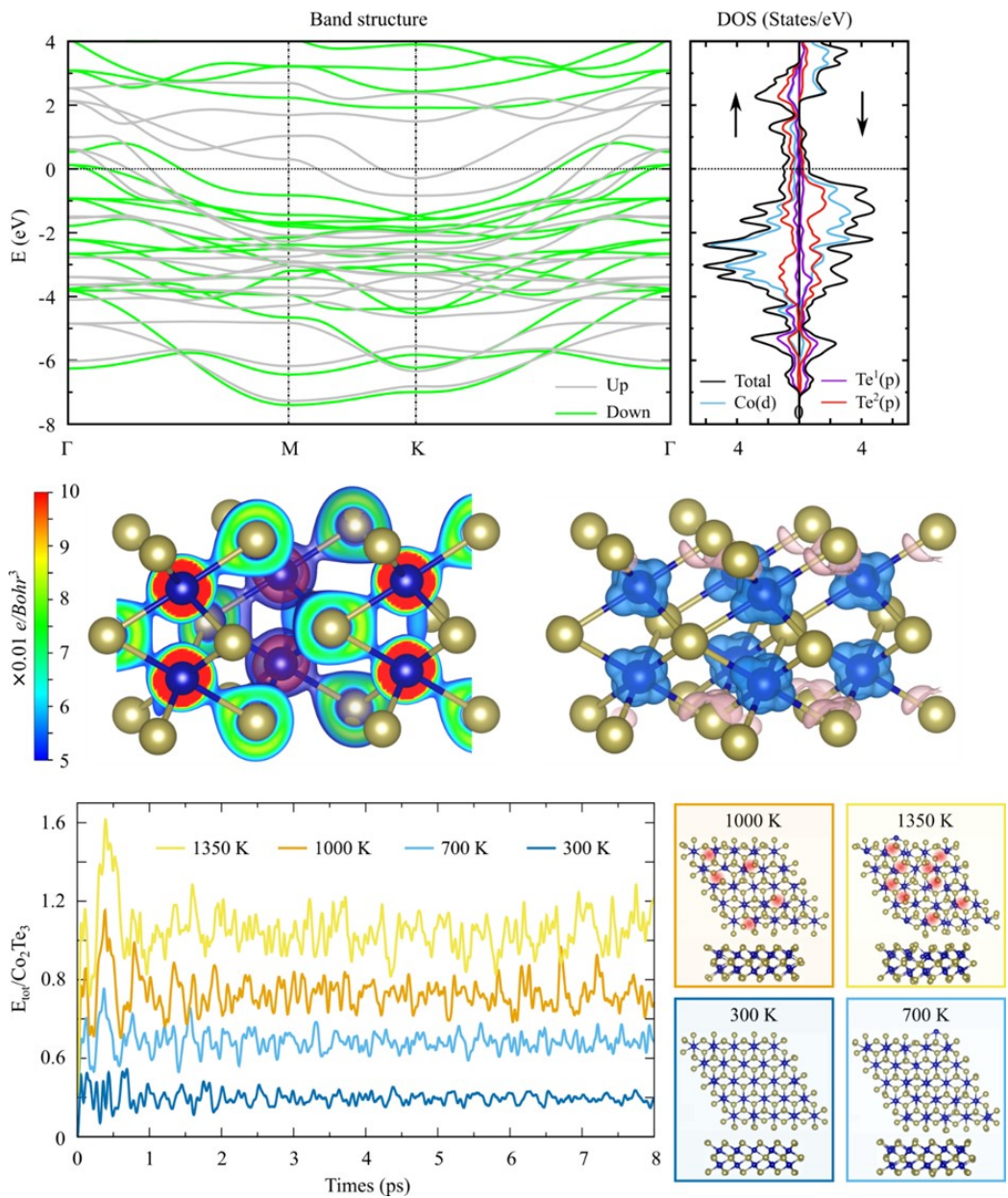

**Figure. 5**: Upper panels are band structure and electronic density of states (HSE06). Mid panels are charge density (red and blue isosurfaces are 0.01 e/Bohr$^3$ and 0.05 e/Bohr$^3$ respectively) and spin densities (blue and pink isosurfaces are 0.0045 e/Bohr$^3$ and -0.0045 e/Bohr$^3$ respectively). Low panels are AIMD simulation results: total energy per formula unit as a function of time; final structures of Co$_2$Te$_3$ monolayer, where broken Co-Te bonds indicated by light-red spots.